\documentclass[authoryear]{elsarticle} 

\usepackage[latin1]{inputenc}
\usepackage{rotating}
\usepackage{graphicx}
\usepackage{epsfig}
\usepackage{subfigure}
\usepackage{fancyhdr}
\usepackage{placeins}

\usepackage{upquote}

\usepackage[left=3cm,right=3cm,top=2cm,bottom=3cm]{geometry}

\newcommand{\pdfauthor}{F. Fabián Rosales Ortega}
\newcommand{\pdftitle}{PINGSoft: an IDL visualisation and manipulation tool for IFS data}

\usepackage[ps2pdf,pdfpagemode=UseOutlines,breaklinks=true]{hyperref}

\hypersetup{citecolor={blue},linkcolor={blue},urlcolor={blue}}     

\hypersetup{
  colorlinks=true,             
  bookmarksnumbered=true,      
  bookmarksopen=false,         
  pdftitle={\pdftitle},        
  pdfauthor={\pdfauthor},      
  pdfdisplaydoctitle           
}
\usepackage[all]{hypcap}

\def\astrobj{}

\begin{document}

\title{
  {\sc PINGSoft}: an IDL visualisation and manipulation tool \\
  for Integral Field Spectroscopic data
}

\vspace{1cm}

\author[ffro1,ffro2]{F.\ F.\ Rosales-Ortega}
\ead{frosales@cantab.net}

\address[ffro1]{Institute of Astronomy, University of Cambridge, Madingley Road,
   Cambridge CB3~0HA, UK.}

\address[ffro2]{Departamento de Astrofísica Molecular e Infrarroja, IEM-CSIC, C/Serrano
   121, 28006, Madrid, Spain.}

\begin{abstract}
In this article we introduce {\sc PINGSoft}, a set of IDL routines designed
to visualise and manipulate, in an interactive and friendly way, Integral Field
Spectroscopic data. The package is optimised for large databases and a
fast visualisation rendering. Here we describe its major characteristics and
requirements, providing examples and describing its capabilities. 
The {\sc PINGSoft} package is freely available at:
\url{http://www.ast.cam.ac.uk/research/pings}
\end{abstract}


\begin{keyword}
techniques: spectroscopic ; methods: data analysis ; integral field spectroscopy
\end{keyword} 

\maketitle

\section{Introduction}

The data reduction, manipulation and visualisation of Integral Field
Spectroscopy (IFS) observations possess an intrinsic complexity given the nature
of the data and the vast amount of information recorded even in a single
observation. In spite of the obvious advantages of this technique in tackling
known scientific problems, in opening up new lines of research, and the
increasing number of instruments available to astronomers, 2-dimensional (2D)
spectroscopy is a technique that is relatively little used.

Nowadays only few groups around the world are capable of reducing and analysing,
in a systematic and homogeneous way, the huge volume of data generated by IFS
observations, and these groups tend to be involved with a particular instrument,
implying that most 2D data reduction, visualization and analysis packages are
orientated towards and limited to a single instrument, so that experience with
one instrument does not necessarily guarantee the ability to work on another.

The data processing of IFS observations requires several steps before any
science can be extracted, and some of them are particular for specific science
cases. They include the (complex) data reduction, mapping, source extraction,
continuum fitting, emission/absorption line fitting, deconvolution,
cross-correlation, etc. Most of these operations are not independent tasks,
and visual/spatial checking of the cube at every step is essential for a
correct data treatment and eventual scientific interpretation.
On this regard, exploring, visualising and manipulating IFS reduced data by
standard means still represents a challenge in the 2D spectroscopy
community, fact that perhaps has discouraged part of the community into
getting involved to this field.

There have been important efforts towards a standardisation of IFS
visualisation and manipulation tools, one is represented by the Euro3D Research
Training Network \citep{Walsh:2002p3819}, who developed the E3D visualization tool
\citep{Sanchez:2004p2632}, a package which allows the user to view an
2D image at any wavelength slice, to explore the spectra at any spaxel, and to
perform simple analyses, being capable of handling several data formats. Its main
limitation resides in the multiple library dependencies during installation,
specially important for non-native linux platforms. {\sc QFitsView} by Thomas
Ott, is a generic FITS viewer program capable of handling IFS data and
performing basic analysis operations on practically any OS flavour. However on
most cases, the final reduced cubes need to be reformatted in order to use this
tool.

Other major astronomical packages include visualisation and manipulation tools
for IFS data as part of their releases, examples are the {\sc GAIA} and 
{\sc DATACUBE} packages, as part of the Starlink project\footnote{See: \url{http://starlink.jach.hawaii.edu/starlink}},
the {\tt wavextract} IRAF\footnote{IRAF is
  distributed by the National Optical Astronomy Observatories, which are
  operated by the Association of Universities for Research in Astronomy, Inc.,
  under cooperative agreement with the National Science Foundation.}
task (by Jeremy Walsh) and some expected 3D tools in the upgrade of the
Groningen Image Processing System, {\sc GIPSY} \citep{vanderHulst:1992p3831}. 
The recently launched {\sc p3d} software for IFS data reduction \citep{Sandin:2010p3798}
also integrates some level of visualisation, but it is restricted to the reduction
pipeline\footnote{More information on IFS software can found at the {\sc IFS wiki}
\citep[][\url{http://ifs.wikidot.com/}]{Westmoquette:2009p3557}, which is a
dedicated webpage with excellent information on nearly every area of the IFS
technique, especially for the novice user.}. 
All these tools allow a quick-look and basic manipulation of IFS observations,
but depend to a certain extent on their major software packages, and in some
cases, on compilation libraries and data formats. Furthermore, they are
restricted to a graphical user interface (GUI), without a command-line based
option, which might be a restriction while handling large IFS databases.

In order to overcome some of these limitations, the {\sc PINGSoft} package was
envisaged, i.e. a set of tools especially designed to visualise and manipulate,
in an interactive and friendly way, IFS data regardless of the original
instrument and spaxel size/shape, able to run on practically any
computer platform and with minimal library requirements. In this article, the
{\sc PINGSoft} package is presented, including a general description of the
program, installation requirements, examples of its performance with real data
and a summary of all routines included within the distribution, in the hope
that the community will find it useful, and in a way to contribute with the
proliferation of IFS-based astronomical research.

\section{The {\sc PINGSoft} package}

The {\sc PINGS software}, or {\sc PINGSoft}, was originally developed during the 
PPAK IFS Nearby Galaxies Survey, or {\sc PINGS} project
\citep{RosalesOrtega:2010p3794}, which used the PMAS
\citep{Roth:2005p2463} Integral Field Unit (IFU) in the PPAK mode
\citep{Verheijen:2004p2481,Kelz:2006p3341,Kelz:2006p338} at the Centro
Astronómico Hispano Alemán (CAHA) at Calar Alto, Spain. Given the large size of
the 2D spectroscopic mosaics observed by PINGS, these routines were conceived to
handle a large amount of data, therefore their implementation for equal or
smaller fields of view and/or with similar instrumental setups is
straightforward.

{\sc PINGSoft} is written in the IDL programming language\footnote{IDL, the
  Interactive Data Language, is a computing environment for data
analysis, data visualization, and software application development, available
from ITT Visual Information Solutions
(\url{http://www.ittvis.com/ProductServices/IDL.aspx}).} 
and consists of a set of individual routines called as command
lines with a specific syntax in an IDL running terminal. The main drawback of
the IDL environment is its commercial character\footnote{Up to date, {\sc
    PINGSoft} cannot be run natively on the GNU Data
  Language, or {\sc GDL}, i.e. the free IDL compatible incremental compiler,
  since the built-in IDL {\tt POLYFILL} procedure is still not available in the
  last GDL release, which is one of the main procedure on which {\sc PINGSoft}
  relies. Once GDL includes this function, the {\em commercial character} of IDL
  preventing the use of {\sc PINGSoft} will not be an issue.}. 
However, nowadays practically
any astronomical centre around the world has a running institutional license of
IDL available for its community, making possible to install very easily any
IDL-based software, without compilation/library dependencies and/or platform
issues.

The {\sc PINGSoft} routines should be seen as the starting point to visualise
and manipulate the (usually) large data formats produced by any IFU
observation. More sophisticated visualisation and analysis would require
tailored-built codes for the specific instrument and scientific case. However,
given that the {\sc PINGSoft} codes are not in the form of pre-compiled
binaries and the source is completely open, the user can extract and modify any
of the routines for more specific and personalised tasks, which is another
advantage compared to other pre-compiled visualisation software.

In terms of data formats, practically any IFS data can be adapted to work with
{\sc PINGSoft}, regardless of the original data format (e.g. 3D cubes, RSS, FITS
tables, etc.), and the size/shape of the spaxel. Given that these routines were
developed for the PPAK instrument, they are perfectly suited for they immediate
implementation to the {\sc CALIFA} data\footnote{Calar Alto Legacy Integral
  Field spectroscopy Area survey, S\'anchez et al. (in preparation) see:\\
  \url{http://www.caha.es/sanchez/legacy/oa/}}.

The public version of {\sc PINGSoft} includes the basic tools to visualise
spatially and spectrally the IFS data, to extract regions of interest by hand or
within a given geometric aperture, to integrate the spectra within a given
region, to read, edit and write IFS FITS files, and to perform simple analyses
to the IFS data. Additionally, some miscellaneous codes useful for generic
tasks performed in astronomy and spectroscopy are also included.
For more sophisticated visualisation/analyses the updated versions
of the {\sc E3D} and {\sc R3D} software
\citep{Sanchez:2004p2632,Sanchez:2006p331} are recommended, which include
several routines for a detailed semi-automated spectroscopic analysis (SSP
continuum fitting, emission line fitting, etc.).

\section{Installation requirements}

All the installation requirements and instructions are explained in detailed in
the {\sc PINGSoft} documentation found at the project webpage. Here we just
summarise the main requirements and installations steps to give the reader an
idea of the {\em complexity} in the installation of {\sc PINGSoft}.

In order to run properly, {\sc PINGSoft} should be installed in a
UNIX, Linux, Mac or Windows computer via a terminal window running any IDL version
greater than 6.0. The user should download the TAR file containing the 
{\sc PINGSoft} library (i.e. {\tt pingsoft.tar.gz}) from the project webpage,
then extract it into a folder of his preference. This will create a directory
named {\tt pingsoft/}, with all the codes of the distribution and additional
subdirectories. In order to work properly, {\sc PINGSoft} requires the entire
content of both the {\sc NASA IDL Astronomy User's Library} and the set of
routines created by David W. Fanning, known as the {\sc Coyote Library}. As
these are very common IDL routines, they are probably already installed as
part of the IDL library of the user's institute. Otherwise, a personal copy
can be downloaded from \url{http://idlastro.gsfc.nasa.gov} and 
\url{http://www.dfanning.com/documents/programs.html} respectively; both
libraries must be extracted and installed in the same way as {\sc PINGSoft}.
Finally, the {\tt pingsoft/} directory (as well as any other new library) and
a {\tt PINGSOFT\_PATH} variable should be defined in the user's system,
e.g. at the startup script that controls the shell of the OS system.

This is everything that the user needs to install the package, to check if the
installation was done correctly, open a {\bf new} terminal and type: 

{\small
\begin{verbatim}
% echo $PINGSOFT_PATH
\end{verbatim}
}

\noindent which should print something like: {\tt /path\_to\_your\_IDL\_directory/pingsoft}\\

\noindent Then, in an IDL running terminal:

\vspace{0.1cm}
{\small
\begin{verbatim}
IDL> check_pingsoft
% Compiled module: CHECK_PINGSOFT.
% Compiled module: PINGSOFT_DISPLAY.

 Checking display...

 Checking paths...

 Checking files...

 PINGSoft installed successfully...!
\end{verbatim}
}

\section{The RSS data format}

{\sc PINGSoft} works with the Row Stacked Spectra or RSS format
\citep{Sanchez:2004p2632,Sanchez:2006p331},
plus a corresponding position table in {\tt ASCII} format. A RSS file consists in a
2D FITS image in which the $X$-axis corresponds to the dispersion axis, and
the other one corresponds to a given spatial ordering of the spectra
determined by the position table, i.e. the $N$-row in the
$Y$-axis corresponds to the spectrum at the position $(X_N,\,Y_N) \equiv
(\Delta {\rm RA}_N,\,\Delta {\rm Dec}_N)$ from a $(0,\,0)$ reference point (in arcseconds),
which is the $N$ entry of the {\tt ASCII} file position table. The number of
rows on the RSS FITS file is equal to the total number of spectra of the IFS
data, i.e. the spectra are {\em stacked one on top of each other} in the
$Y$-axis. This is the standard output format for reduced PMAS/PPAK data, and
for the PINGS and CALIFA projects.

For instruments in which the standard output is a 3D spectral cube, individual
spectra can be extracted at any spatial position and stored in a RSS file. The
spatial location of the spectra can be then recorded in a {\tt ASCII} file to create
the position table. This technique has been tested successfully with data
obtained with the VIMOS instrument \citep{LeFevre:1998p3053}. The {\tt
  cube2rss.pro} routine included in {\sc PINGSoft} can easily convert a 3D cube into a
RSS file.\\

\noindent The format of the input position table should be the following:

\vspace{0.2cm}
{\small
\begin{verbatim}
C       1.34       1.34      1
1     -13.936      0.000     1
2      15.686      3.019     1
3       0.000     12.071     1
4     -13.936     12.077     1
5      13.935     12.077     1
6     -10.452      6.038     1
...
\end{verbatim}
}

\noindent where the first line determines the size and shape of the spaxel
(see below), and the following entries correspond to:
{\tt  ID$_{\#}$     X$_{\rm offset}$    Y$_{\rm offset}$   flag}, 
where {\tt ID$_{\#}$} is the spectrum number identification (integer
value) and {\tt flag} might be any numerical value (used sometimes for
internal quality control).
The size and shape of the spaxels is determined by the {\em first two entries}
of the first line of the position table, the first character should be either a
{\tt C} or {\tt S}, which corresponds to a {\bf c}ircular (e.g. fibre, PPAK)
or {\bf s}quare (e.g. PMAS, VIMOS) spaxel, and the second entry should be a
floating number corresponding to the radius or side length respectively, e.g.:\\

{\footnotesize
\begin{verbatim}
    C  1.34    ('C'ircular spaxel, with a radius 1.34 arcsec, e.g. PPAK)

    S  0.67    ('S'quare spaxel, with sides of length 0.67 arcsec, e.g. VIMOS)
\end{verbatim}
}
\vspace{0.3cm}

\noindent the {\tt \$PINGSOFT\_PATH/pos\_tables} directory contains more
position table examples.\\

In addition to the format restrictions mentioned above, the RSS file should be
wavelength calibrated and the spectral information
should be included in the FITS header, namely the {\tt CRPIX1}, {\tt CRVAL1}
and {\tt CDELT1} values. If the IFS data to analyse is in the RSS format with
corresponding position tables, then {\sc PINGSoft} should work
smoothly and the visualisation/manipulation should be straightforward (e.g. {\sc
CALIFA} data).

\section{{\sc PINGSoft} by examples}

All the {\sc PINGSoft} routines are called via command lines in
a terminal running IDL. The syntax of any program can be obtained by
entering the name of the procedure without any other parameter. The routines
are also documented within the {\tt .pro} files, where an explanation, syntax
and examples of the routines are included as comments at the beginning of the
file.

The potential of the software can be illustrated by introducing the
prototype procedure from which the rest of the main routines were based. This
code is called {\tt view\_rss.pro}, which displays interactively a
visualisation of any RSS file, including its spatial and spectral information. 
The {\sc PINGSoft} distribution includes some RSS example files in the directory 
{\tt \$PINGSOFT\_PATH/examples/}, they correspond to different versions of
a PINGS dithered observation of the central part of \astrobj{NGC~4625}, covering from
6000 to 6650 \AA; the main files are the following:

\begin{itemize}
\item {\tt pings.n4625\_331.fits}: single PPAK exposure, corresponding to the
  central hexagon of the instrument; the RSS file contains 331 spectra.
\item {\tt pings.n4625\_382.fits}: same as before, but including the sky and
  calibration fibres of the instrument, for a total of 382 spectra.
\item {\tt pings.n4625\_dither.fits}: mosaic of the dithered observation including
  the three exposures (without the sky/calibration fibres); the RSS file
  contains 993 spectra.
\item {\tt pings.n4625\_pos\#.fits}: individual pointings of the mosaic
  described above, where \# goes from 1 to 3; each RSS file contains 331
  spectra.
\end{itemize}

\noindent These files will be used in the following sections in order to
introduce the main routines of {\sc PINGSoft}. All the example commands used
in this article can be found in the file {\tt PINGSoft\_examples.pro}.

\subsection*{\Large \tt view\_rss}
\label{view_rss}
\vspace{0.2cm}

This routine provides a 2D interactive visualisation of the spaxels and
spectra of a RSS file.
Generally the program requires the name (and path) of the RSS file and its
position table (as IDL strings). However, there are a number of special
cases in which the program identifies the format of the RSS file (by the
number of spectra and header information), and therefore the user does not need
to include the entry for the position table, these cases are:

\vspace{0.3cm}
\begin{enumerate}
\item PMAS single pointing, all three resolutions ($16\times16$ spaxels).
\item PPAK single pointing (331 spectra).
\item PPAK 3 pointings dither mosaic (993 spectra).
\item Full PPAK pointing, including the sky and calibration fibres (382 spectra).
\item VIMOS single pointing, all resolutions in both configurations$^1$
  ($40\times40$ and $80\times80$ spaxels).
\item VIMOS HR dithered `super-cube' (square 4 pointing dither
  pattern, $44\times44$ spaxels, $29.7''\times29.7''$).
\end{enumerate}
\vspace{0.3cm}

\noindent For example, to visualise a single PPAK pointing of \astrobj{NGC~4625}
(central hexagon, 331 fibres), type:

\vspace{0.2cm}
{\small
\begin{verbatim}
IDL> view_rss, 'pings.n4625_331.fits'
\end{verbatim}
}
\vspace{0.2cm}

\noindent which produces the same visualisation as:

\vspace{0.2cm}
{\small
\begin{verbatim}
IDL> view_rss, 'pings.n4625_331.fits', 'ppak_331.txt'
\end{verbatim}
}
\vspace{0.2cm}

\noindent i.e. including explicitly the name of the position table.\\

\begin{figure}[!t]
  \begin{center}
    \includegraphics[height=5cm]{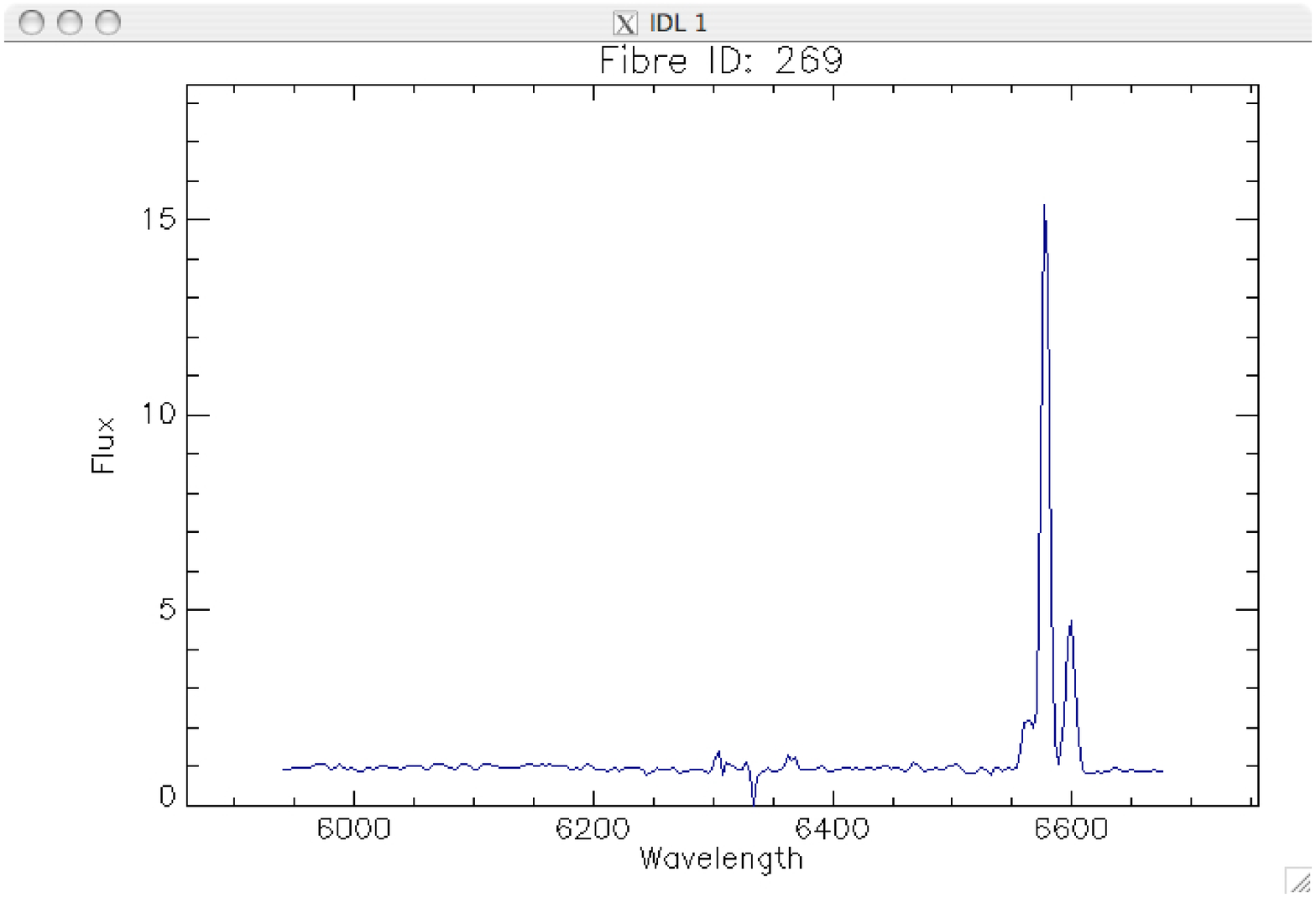}\hspace{0.3cm}
    \includegraphics[height=5cm]{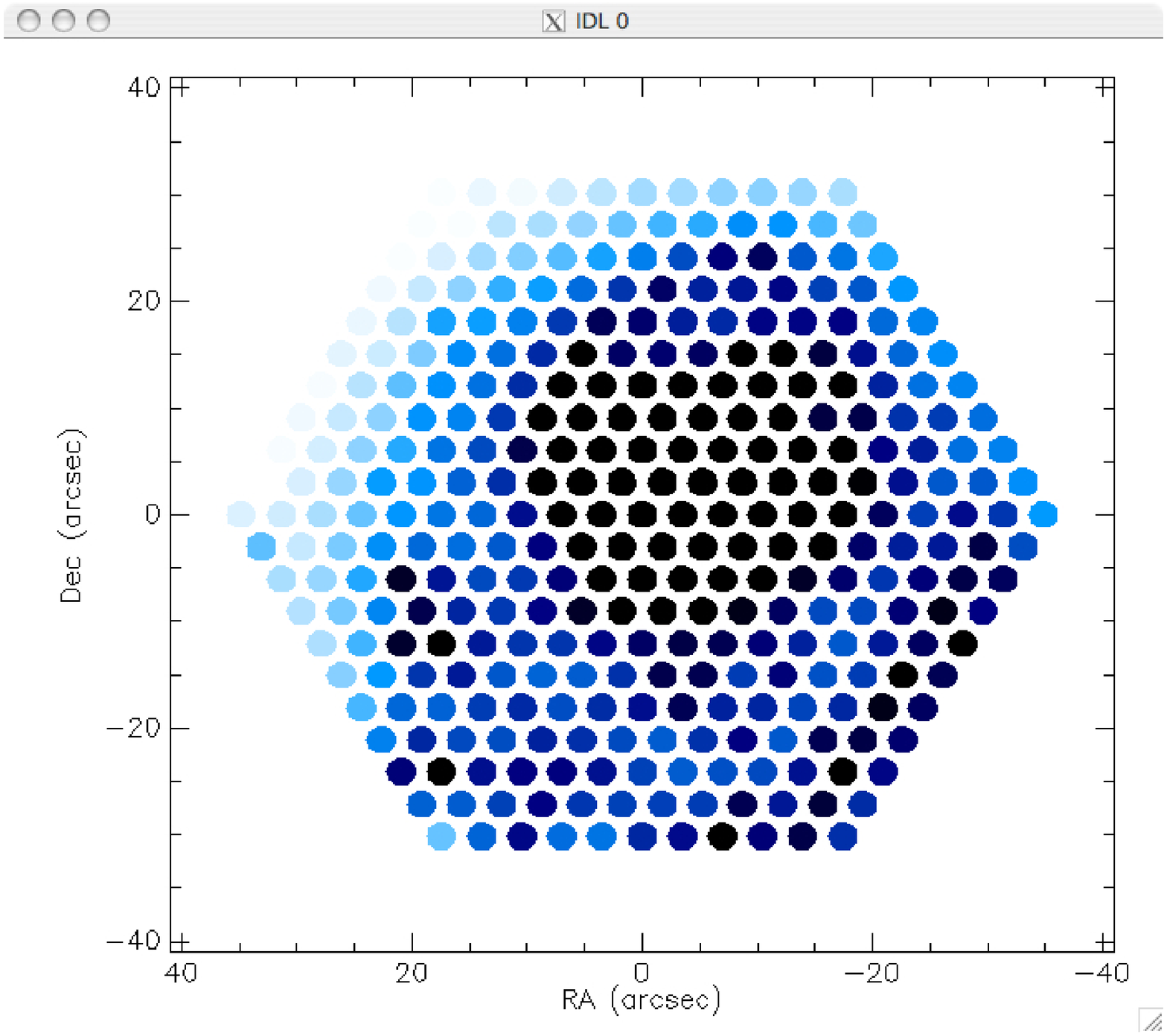}
    \caption[]
    {Screen shots of the visualisation windows generated by the {\tt
    view\_rss} command, on the right the spatial distribution of spaxels, on
    the left the spectral window showing the spectrum of a particular region
    of \astrobj{NGC~4625}.
    \label{fig:view_rss}
    }
  \end{center}
\end{figure}

\noindent The program displays two windows (see \autoref{fig:view_rss}), on the right a
visualisation of the spatial distribution of spaxels. The color-scale
corresponds to a pseudo-narrow band image of a certain width centered at a
given wavelength\footnote{Default width: 100 \AA. Default wavelength:
  H$\alpha$, $\lambda$6563 if within the spectral range, otherwise is equal to
  the mean of the wavelength range.}. The spatial units are assumed to be arcseconds in a
standard North (up) East (left) configuration.
On the left, the window shows the spectrum of the spaxel corresponding
to the position of the mouse, the wavelength range is extracted from the
information on the RSS FITS header, the fibre ID shown on the top of the
window corresponds to the position of the spectrum in the RSS file
(in the IDL format, i.e. starting at zero).\\

Additional information is shown on the IDL terminal where the program was
called, a left-click prints the spaxel information including the fibre ID, the
offset from the reference spaxel (in arcsec), and if the WCS information is
included in the FITS header, it shows the coordinates of the
spaxel in sexagesimal and degree units\footnote{Set the width of the terminal big
enough so that you can see correctly this information.}, e.g.\\

{\scriptsize
\begin{verbatim}
IDL> view_rss, 'pings.n4625_331.fits'

 RSS spectra viewer
 ==================

 Move mouse over the mosaic to plot the spectra

 Options:

   LEFT-click: spaxel information

 MIDDLE-click: selects and stores spaxels by subsequent left-clicks

  RIGHT-click: QUIT


  Spaxel ID    RA offset   Dec offset         RA               Dec            RA deg    Dec deg

       269      17.4200     -12.0779     12h 41m 54.6s    41d 16m 10.8s     190.47765  41.269655
       164     -3.48400      0.00000     12h 41m 52.8s    41d 16m 22.8s     190.46992  41.273010
       267     -8.71893     -21.1184     12h 41m 52.3s    41d 16m  1.7s     190.46799  41.267144
        39      10.4520      12.0779     12h 41m 54.0s    41d 16m 34.9s     190.47507  41.276365
\end{verbatim}
}
\vspace{0.3cm}

\noindent A middle-button click prompts for a {\tt PREFIX} used to generate a new series
of files, all the subsequent left-clicks over the interactive right window
will store the position and index of the selected spaxels which are
outlined in red. When the program is terminated (by a right-click on the right
window) the following files are created:

\vspace{0.2cm}
{\scriptsize
\begin{verbatim}
    Extracted RSS:   PREFIX_rss.fits    (Extracted RSS of the selected spaxels)
   Position table:   PREFIX_pt.txt      (Position table of the new RSS file)
      IDL indices:   PREFIX_index.txt   (Original indices of the selected spaxels)
 Integrated ASCII:   PREFIX_integ.txt   (Integrated spectrum in ASCII format)
  Integrated FITS:   PREFIX_integ.fits  (Integrated spectrum in FITS format)
\end{verbatim}
}
\vspace{0.2cm}

\noindent and the spectral window shows the integrated spectrum of the
selected spaxels.\\

\begin{figure}[!t]
  \begin{center}
    \includegraphics[height=7cm]{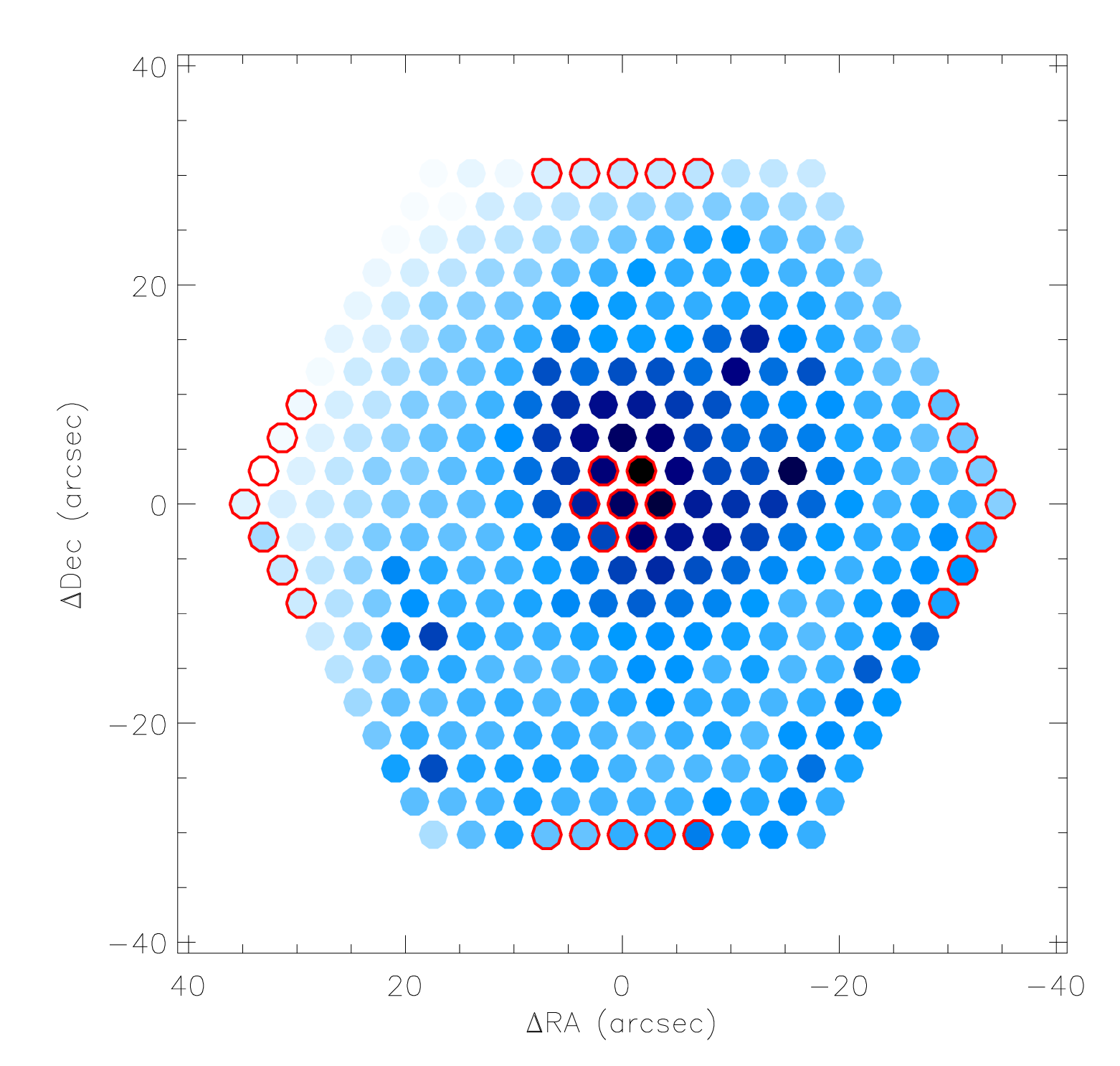}\hspace{0.5cm}
    \includegraphics[height=7cm]{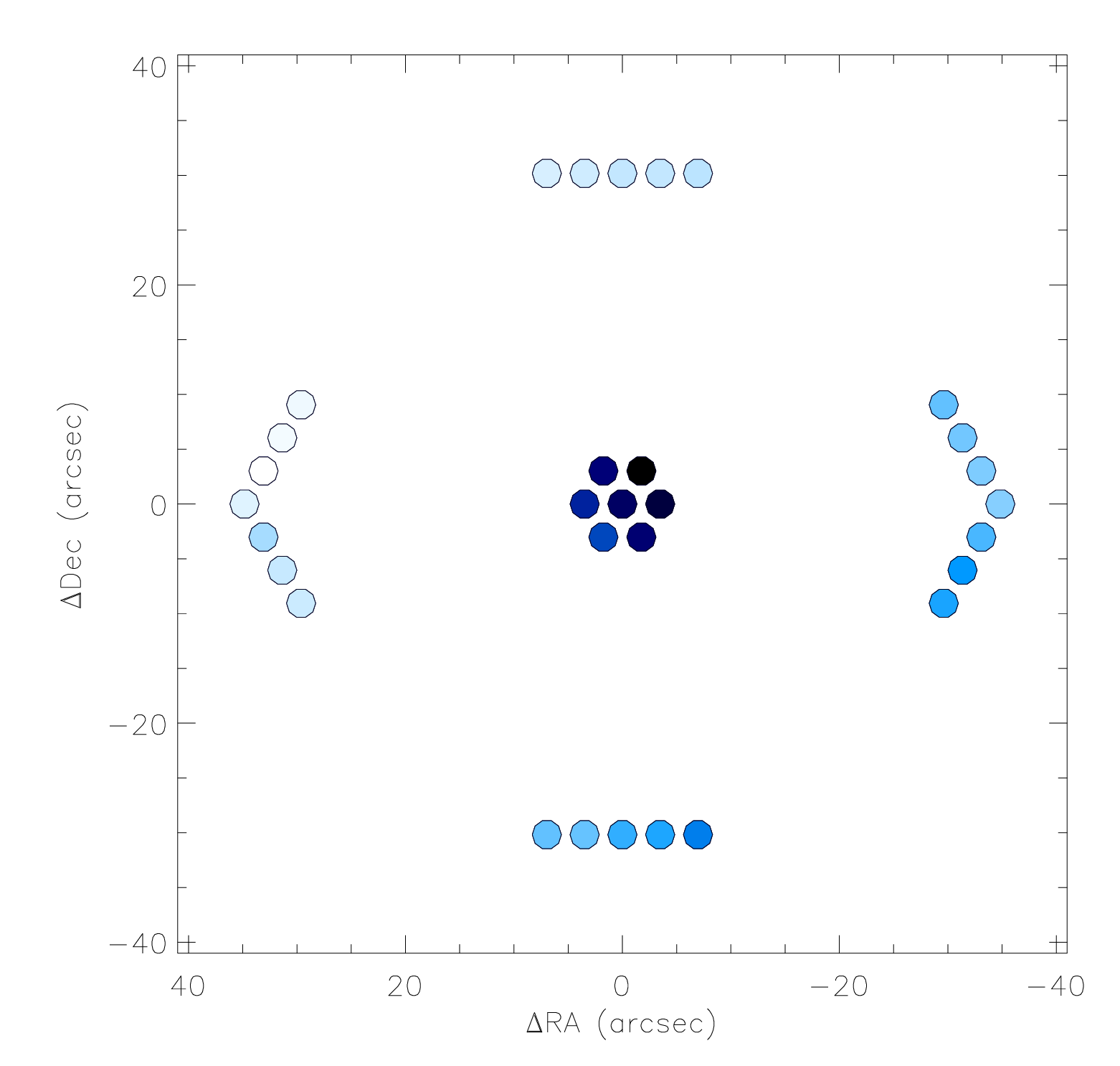}
    \caption[]
    {Visualisation examples of spaxel extraction using the middle-click option of the 
      {\tt view\_rss} command. On the left, a single frame of \astrobj{NGC~4625}
      showing the selected spaxels outlined in red. On the right,
      visualisation of the selected spaxels by using the {\tt view\_rss}
      command on the generated RSS and position table files.
    }
    \label{fig:middle}
  \end{center}
\end{figure}

The left panel of \autoref{fig:middle} shows an example of
some selected spaxels from the previous visualisation. The extracted RSS can
be visualised again using {\tt view\_rss}, e.g. if the chosen prefix was
{\tt test1}, then the selected spaxels can be displayed by:

{\small
\begin{verbatim}
IDL> view_rss, 'test1_rss.fits', 'test1_pt.txt', /draw
\end{verbatim}
}

\noindent where {\tt test1\_rss.fits} is the new created RSS file and {\tt test1\_pt.txt} 
is the corresponding position table file, both shown in the right panel of
\autoref{fig:middle}.\\

Several options are available to better visualise the data, including the
central wavelength, width and intensity scaling of the pseudo narrow-band
image, different fonts and colour tables, flux intensity and spectral ranges,
etc. If the command name of any {\sc PINGSoft} routine is entered without any
keywords, the user will get online help with the correct syntax and available
options, e.g.\\

{\scriptsize 
\begin{verbatim}
IDL> view_rss

 CALLING SEQUENCE: 

        view_rss, 'RSS.fits' [, 'pos_table.txt', MIN_FLUX=min_flux, MAX_FLUX=max_flux, LMIN=lmin, LMAX=lmax, $
                             BAND=band, WIDTH=width, CT=ct, FONT=font, /DRAW, /LINEAR, /GAMMA, /LOG, /ASINH, /PS]


      'RSS.fits':  String of the wavelength calibrated RSS FITS file.

 'pos_table.txt':  String of the position table of the RSS file in ASCII format (in NE configuration).
                   (compulsory if not included in the default instruments/setups)

    MIN/MAX_FLUX:  Minimum/maximum flux in the spectral window to be plot,
                   if not set these are floating value.

       LMIN/LMAX:  Defines wavelength range on the spectral window, 
                   if not set values are taken from the RSS header.

            BAND:  Central wavelength of the narrow band used to display the data, 
                   defaults: BAND=6563 (i.e. Halpha) if within the spectral range, 
                   else BAND=mean(lambda).

           WIDTH:  Width of the band to display the data, if not set the default is 100 Angstroms.

              CT:  IDL Color Table used to display the data (default ct=1, BLUE/WHITE).

            FONT:  Vector-drawn IDL font to be used, default: 3 (Simplex Roman)

           /DRAW:  Draws the contour of the spaxels.

         /LINEAR:  Displays the range of intensities using a linear min/max scaling.

          /GAMMA:  Displays the range of intensities using a power-law (gamma) scaling.

            /LOG:  Displays the range of intensities using a logarithmic scaling.

          /ASINH:  Displays the range of intensities using an inverse hyperbolic sine function scaling.

             /PS:  Writes a Postscript file of the spaxels visualisation.
\end{verbatim}
}
\vspace{0.5cm}

\begin{figure}[!t]
  \begin{center}
    \includegraphics[height=7cm]{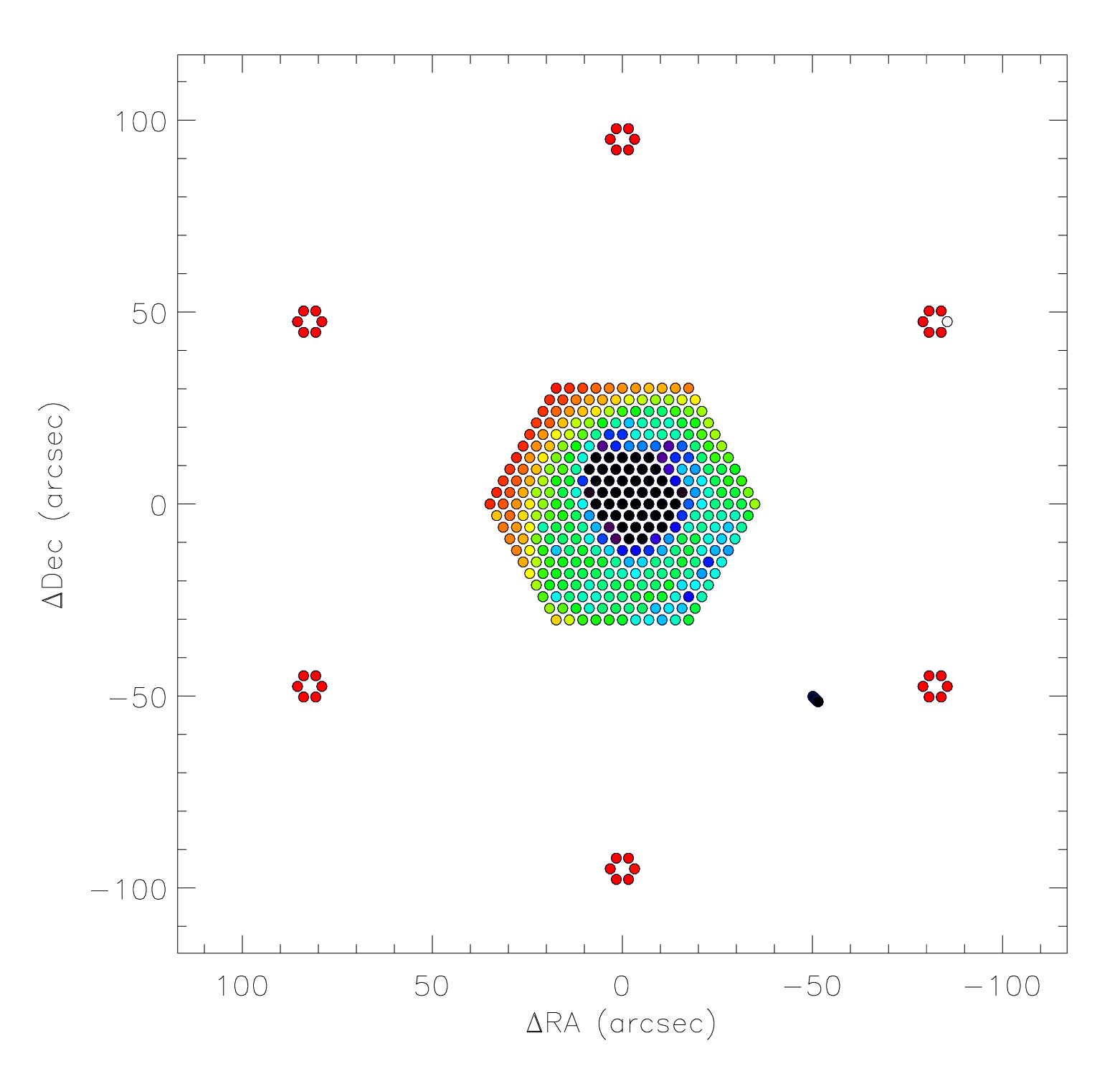}\hspace{0.5cm}
    \includegraphics[height=7cm]{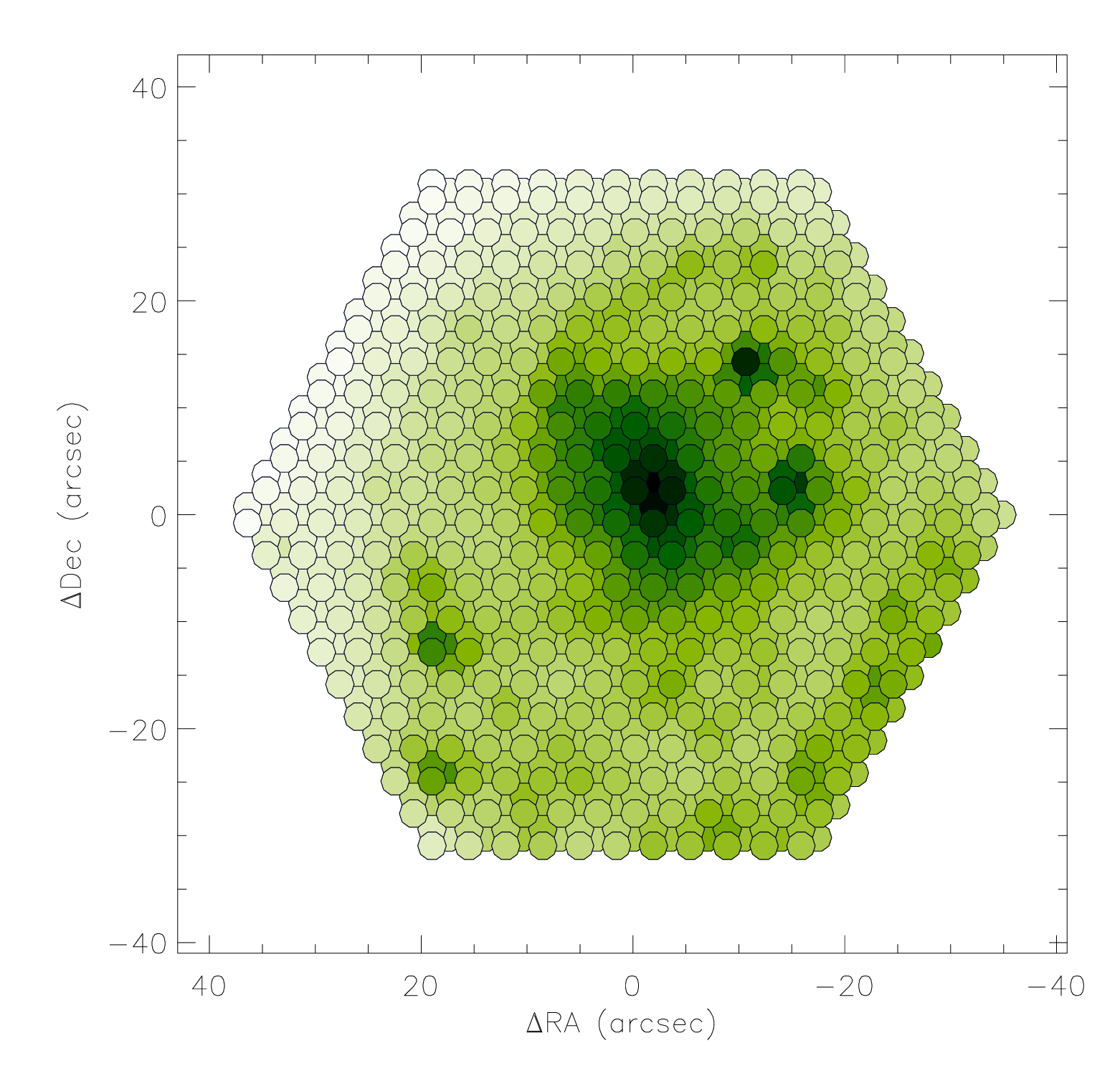}
    \caption[]
    {Example of visualisation windows generated by the {\tt view\_rss} command
    using the {\tt /PS} option. The left panel corresponds to the full bundle
    of the PPAK instrument, including the sky and calibration fibres. The
    right panel corresponds to the dithered mosaic of the nuclear part of
    \astrobj{NGC~4625}. See the text for more details.
  }
    \label{fig:examples}
  \end{center}
\end{figure}

\noindent If the {\tt /PS} keyword is set, the program does not display a
visualisation on the screen, instead writes a Postscript file of the distribution
of spaxels (right window) with the selected display options (see \autoref{fig:examples}).\\
\vspace{0.5cm}

\begin{figure}[!t]
  \begin{center}
    \includegraphics[height=7cm]{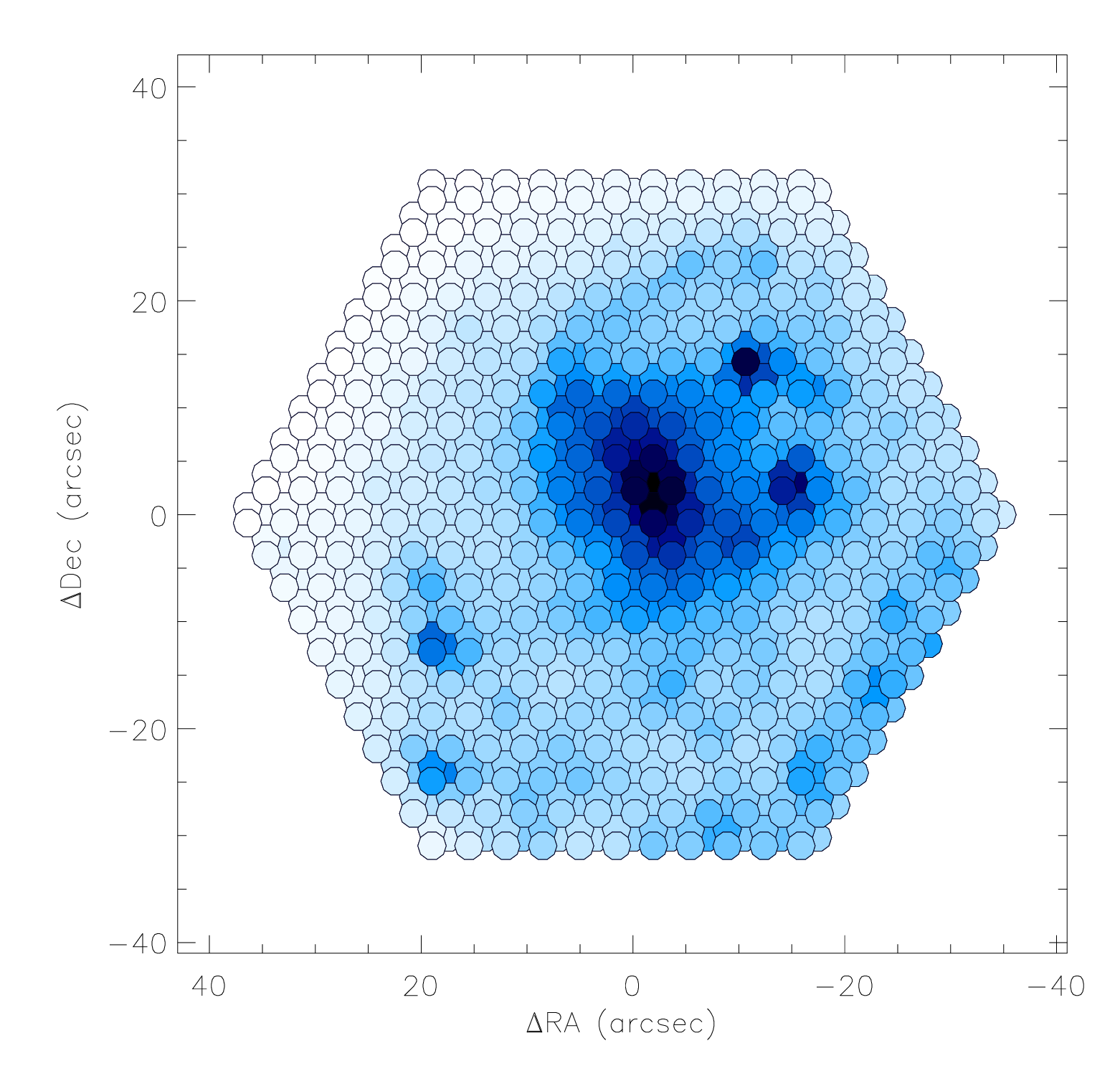}\hspace{0.5cm}
    \includegraphics[height=7cm]{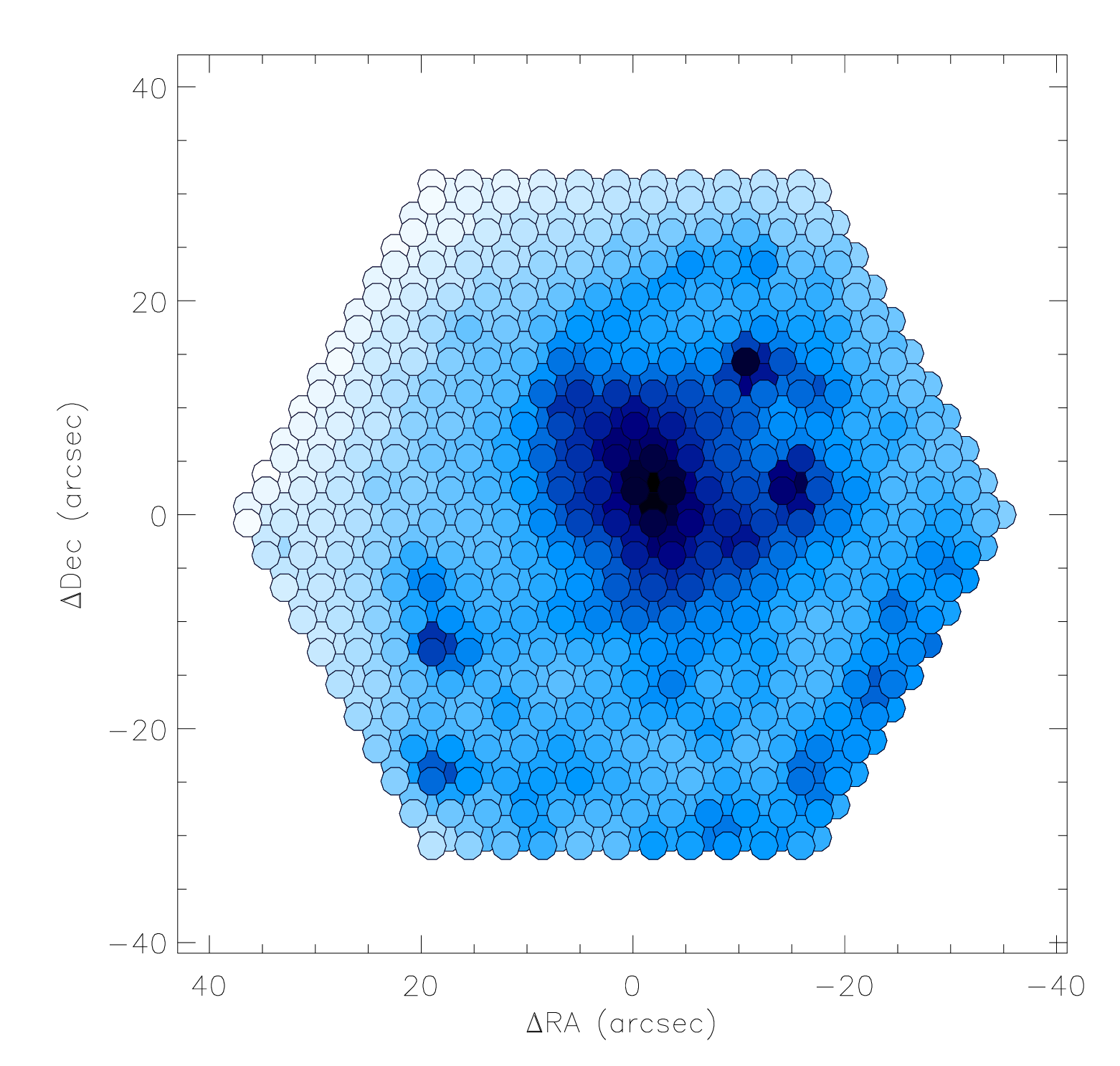}
    \includegraphics[height=7cm]{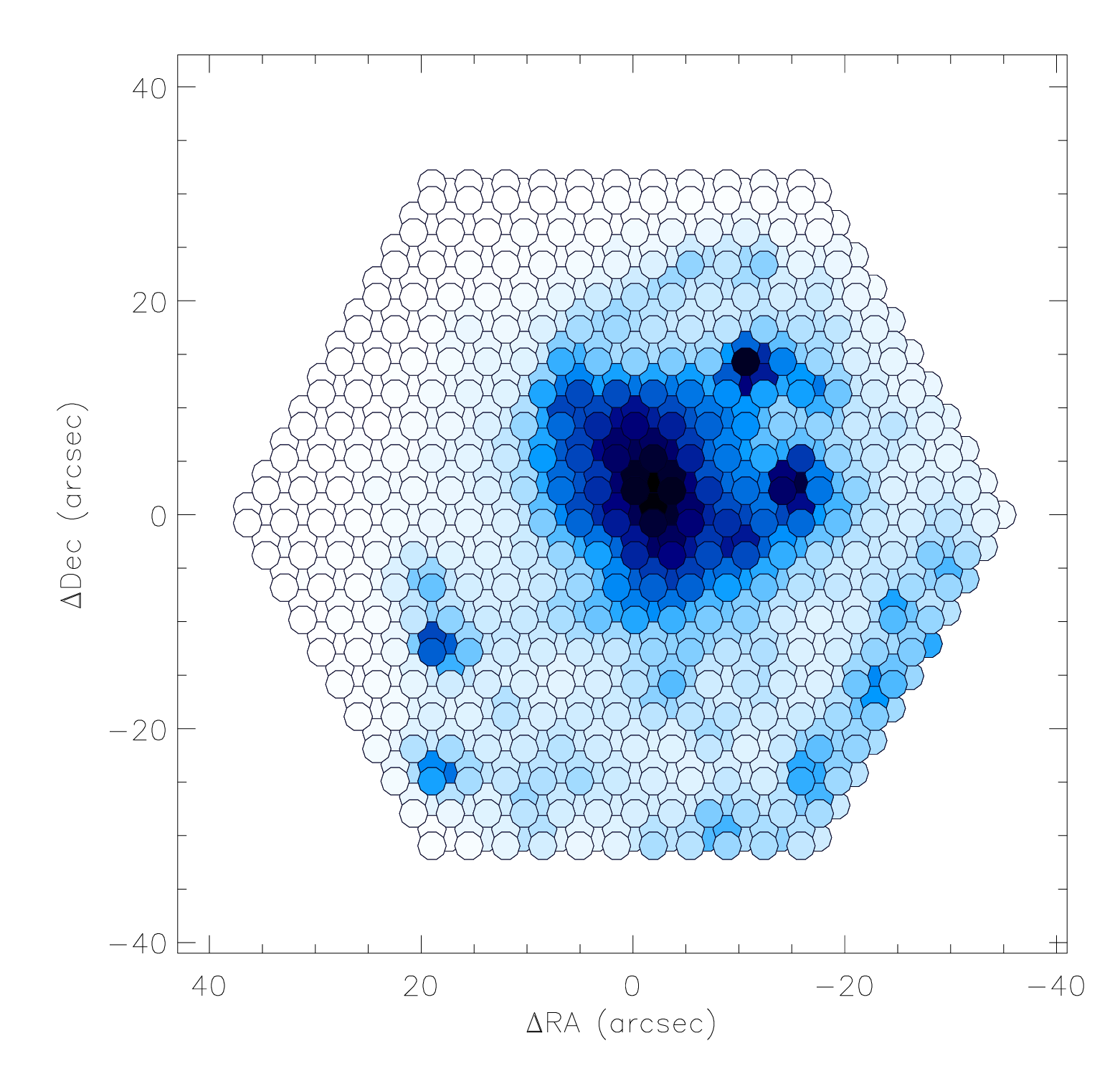}\hspace{0.5cm}
    \includegraphics[height=7cm]{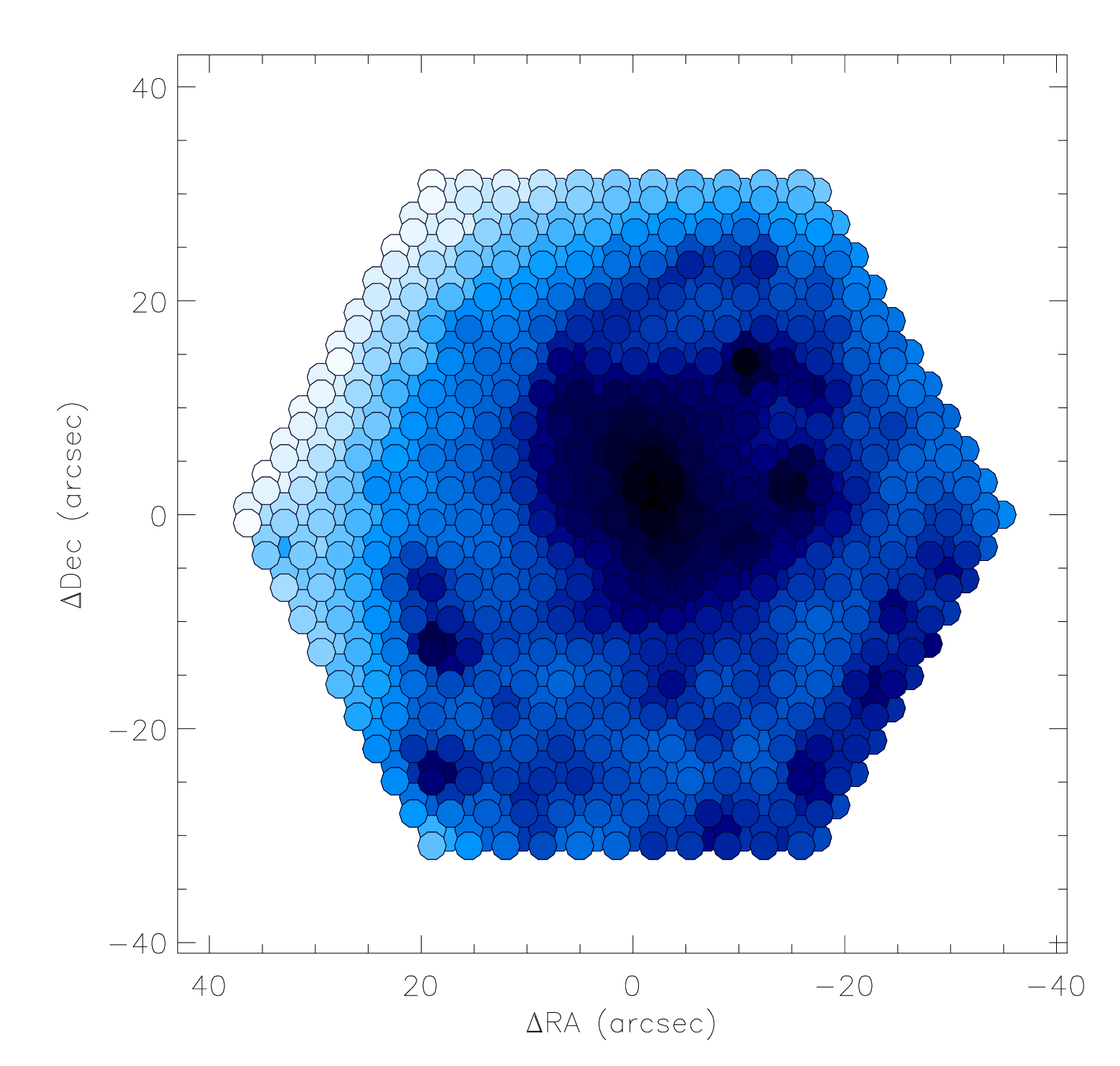}
    \caption[]
    {Examples of different visualisations of \astrobj{NGC~4625} generated by the {\tt view\_rss} command
    using different intensity scaling functions. Top-left panel: a linear scaling, {\tt
    /LINEAR}; top-right: a power-law function, {\tt /GAMMA}; bottom-left: a
    logarithmic scaling, {\tt /LOG}, bottom-right: an inverse hyperbolic sine
    function, {\tt /ASINH}. See the text for more details.
    }
    \label{fig:scaling}
  \end{center}
\end{figure}

\noindent {\sc Intensity scaling}
\vspace{0.4cm}

The default colour-scale of the visualisation is obtained by sampling the
range of intensities within the chosen narrow band into an {\em ad hoc} colour
dynamic range of 255 values (i.e. equal to the number of values in a given IDL
colour table). 
However, very different ranges of intensities are expected depending of the
object, spectral range and signal-to-noise of the observations. Given that the
main purpose of this routine is to visualise easily the IFS data, several
intensity scaling functions are available to the user in order to improve the
contrast and the identification of spectral features: a full linear
(min/max) sampling, a power-law (gamma) function, a logarithmic
transformation, and an inverse hyperbolic sine scaling.

For the {\tt /GAMMA}
transformation, the default value is $\gamma = 0.7$; for the {\tt /LOG}
scaling, the exponent default value is 2; for the {\tt /ASINH} function, the
default $\beta$ value is 10. For a full explanation of the scaling functions
and these parameters see the documentation of the individual routines.
\autoref{fig:scaling} shows the visualisation of the dithered mosaic of
\astrobj{NGC~4625} using the different intensity scalings mentioned above.\\







\begin{figure}[!t]
  \begin{center}
    \includegraphics[height=7cm]{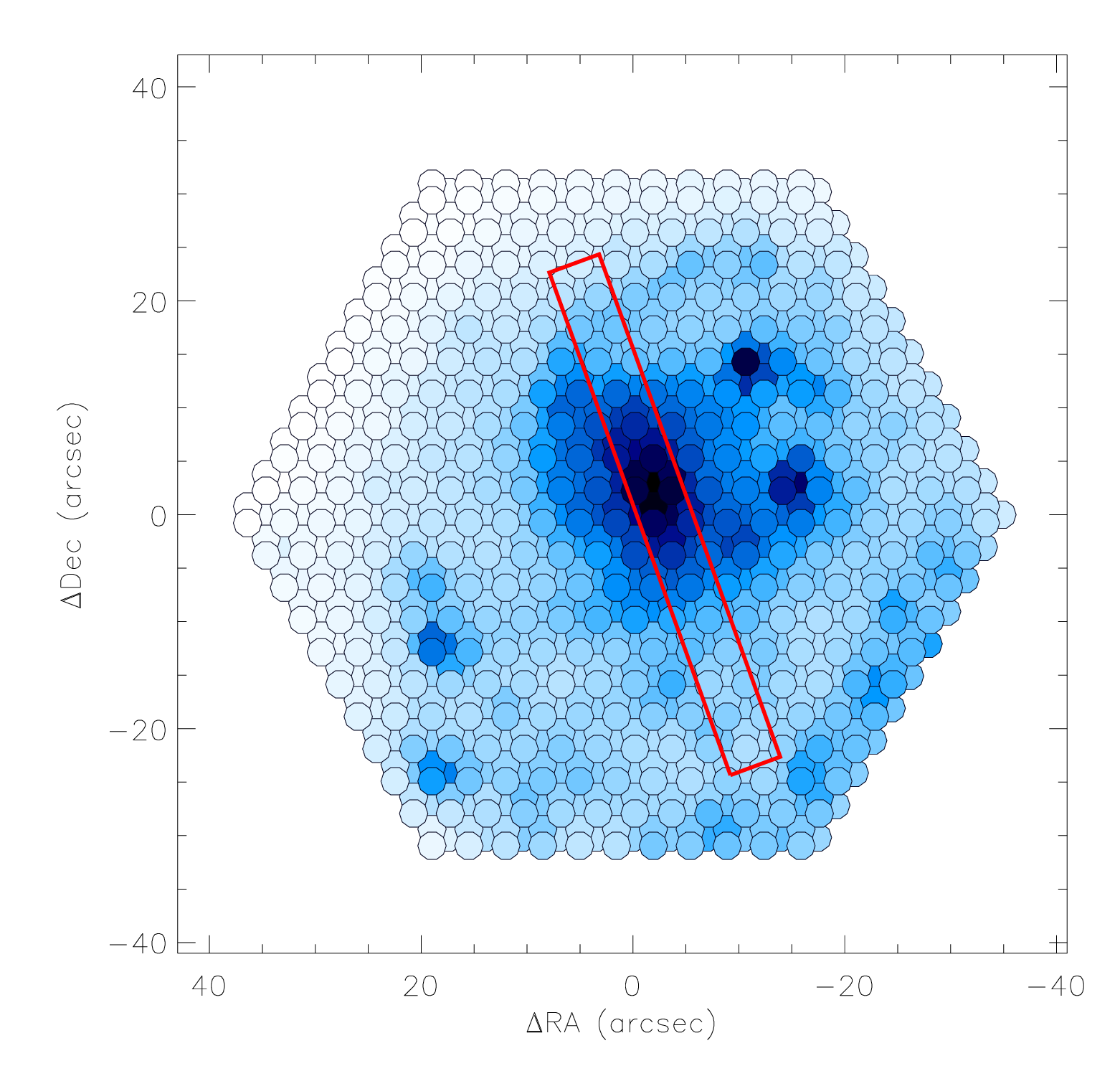}\hspace{0.2cm}
    \includegraphics[height=7cm]{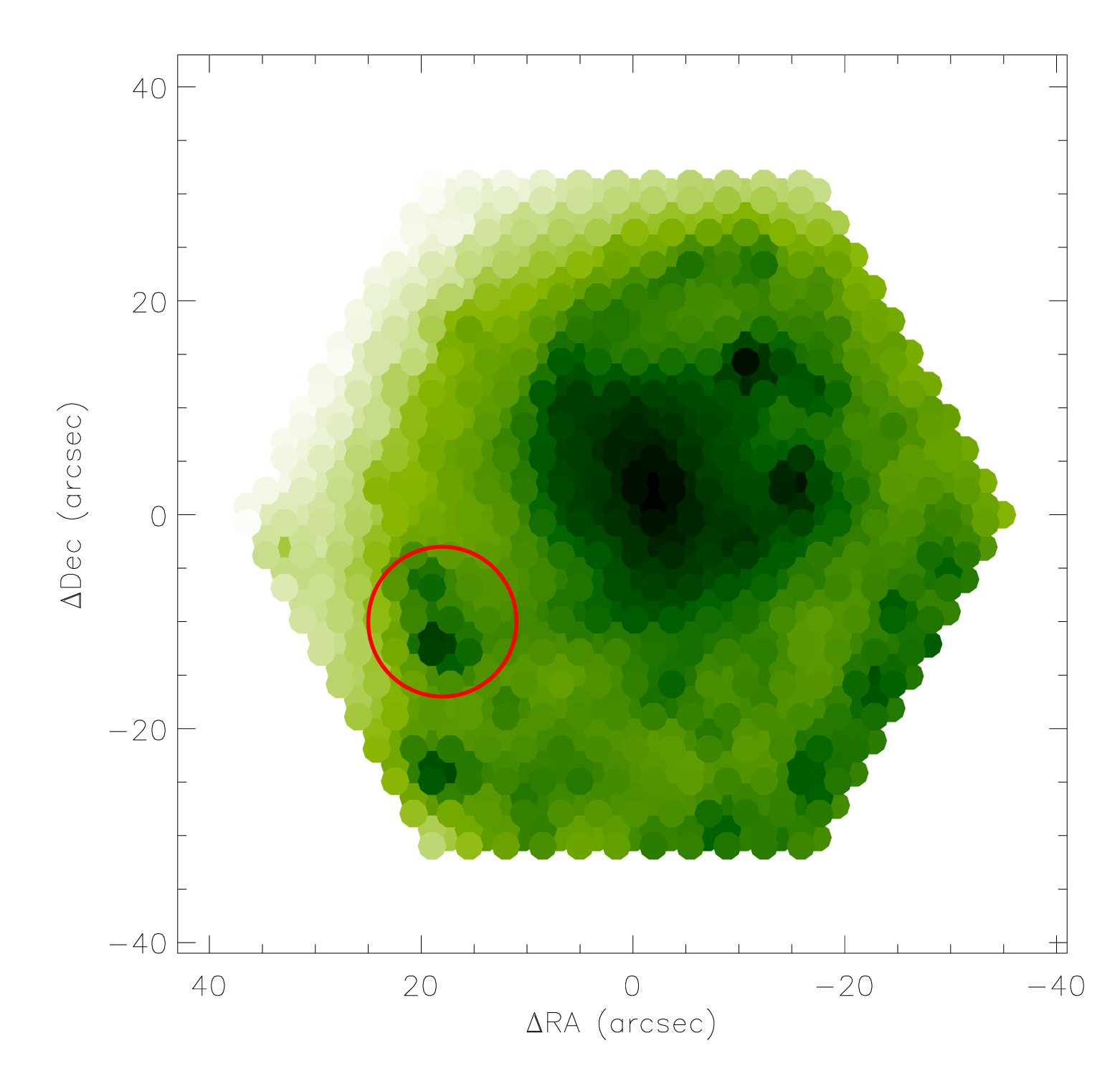}
    \includegraphics[height=6cm]{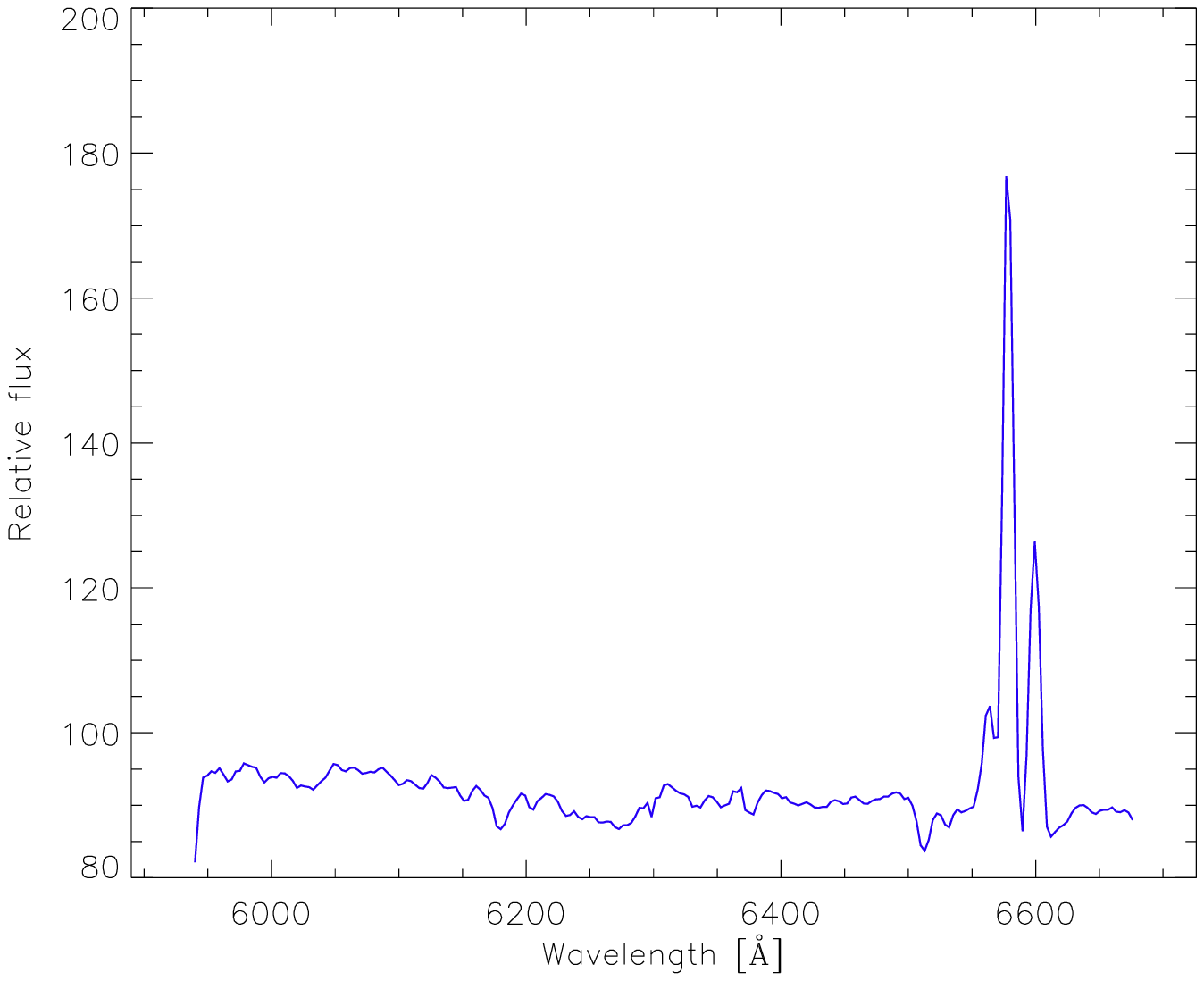}\hspace{0.2cm}
    \includegraphics[height=6cm]{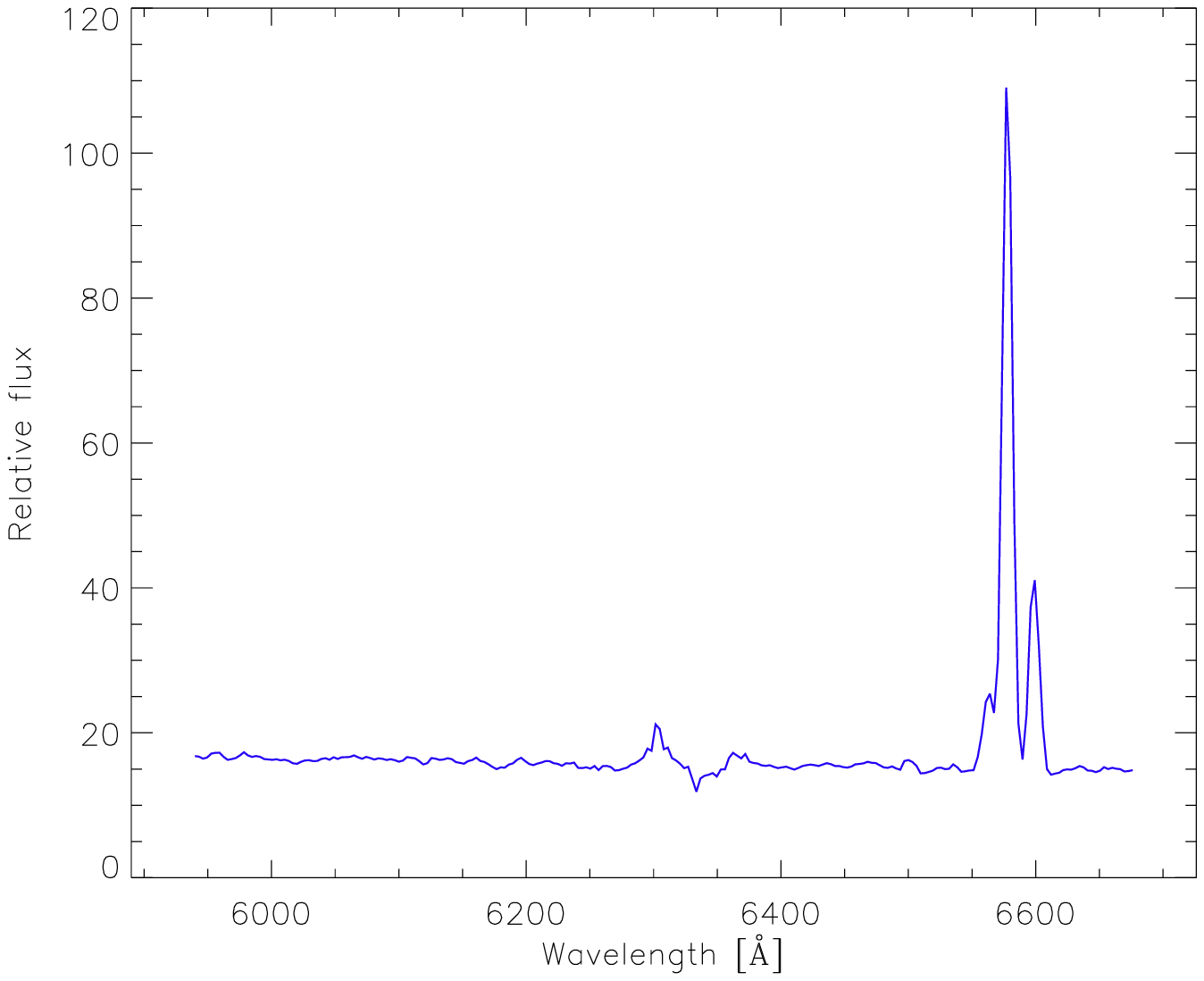}
    \caption[]{
      Examples of spectra extraction using a slit aperture (top-left) and a
      circular aperture (top-right) on the dithered mosaic of \astrobj{NGC~4625}, using the
      {\tt extract\_slit} and {\tt extract\_aperture} respectively. The bottom
      panels show the integrated spectra extracted within each aperture.
    \label{fig:extract}
    }
  \end{center}
\end{figure}

\section{{\sc PINGSoft} list of routines}
\label{sec:list}

The main visualisation code in {\sc PINGSoft} is {\tt view\_rss}, 
there is an additional routine for visualising specific sections of a RSS file
where the ID of the spaxels are known (see below). The rest of the routines in
the {\sc PINGSoft} can be classified into three categories: a) Spectra
extraction and integration; b) RSS and FITS manipulation and c) Miscellaneous
codes. In this section we include a description of all the available {\sc PINGSoft}
routines, it is beyond the scope of this article to explain every single code
and give examples in each case. Detailed descriptions can be found in the
documentation.\\
\vspace{1cm}

\noindent{\bf Visualisation}
\vspace{0.4cm}

\noindent{\tt view\_rss}:
Provides a 2D interactive visualisation of the spaxels and spectra of a RSS file.
\vspace{0.2cm}

\noindent{\tt view\_spaxel}:
Displays the spectra of previously selected spaxels.
\vspace{0.8cm}

\begin{figure}[!t]
  \begin{center}
    \includegraphics[height=7cm]{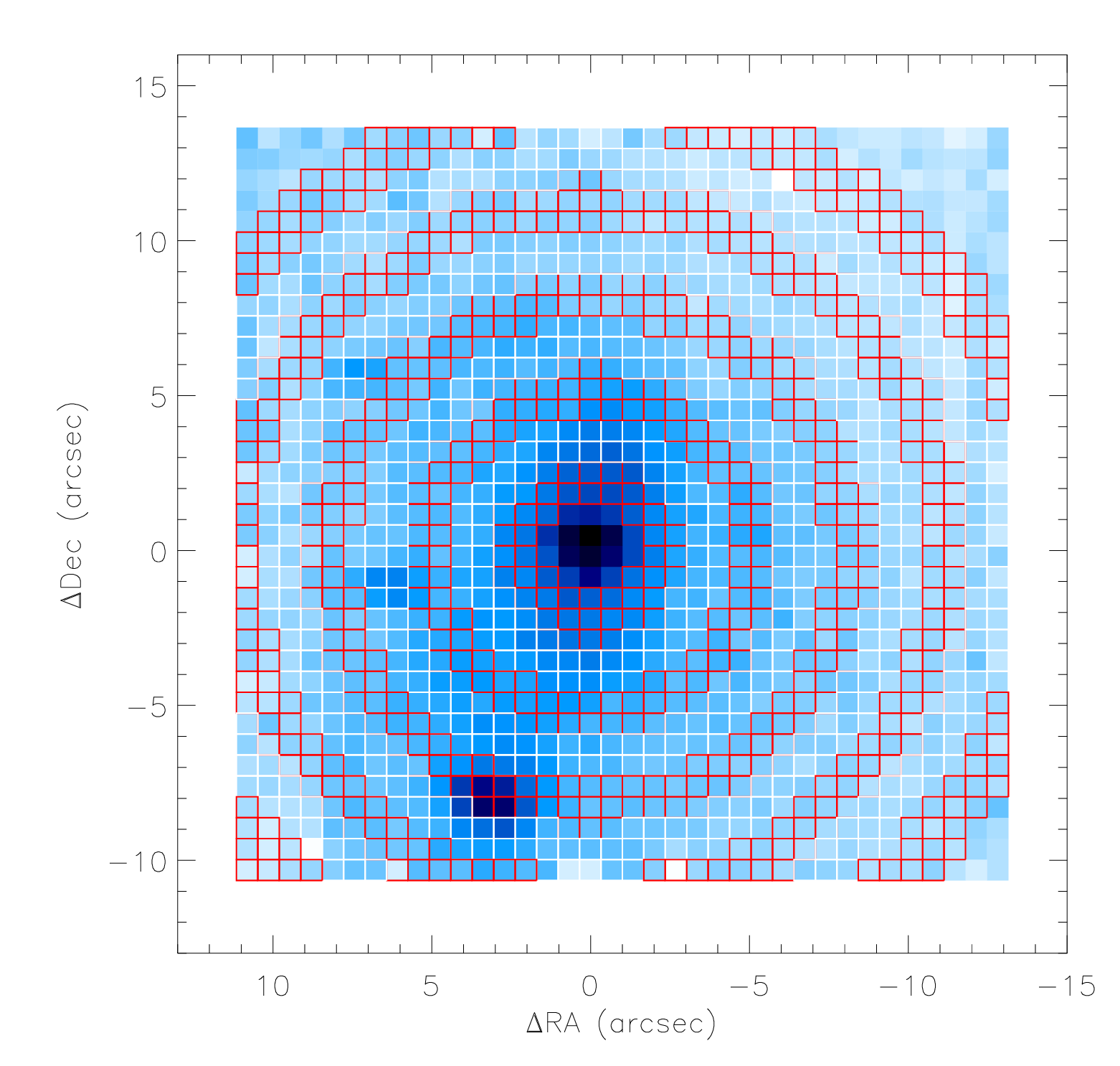}\hspace{0.2cm}
    \includegraphics[height=7cm]{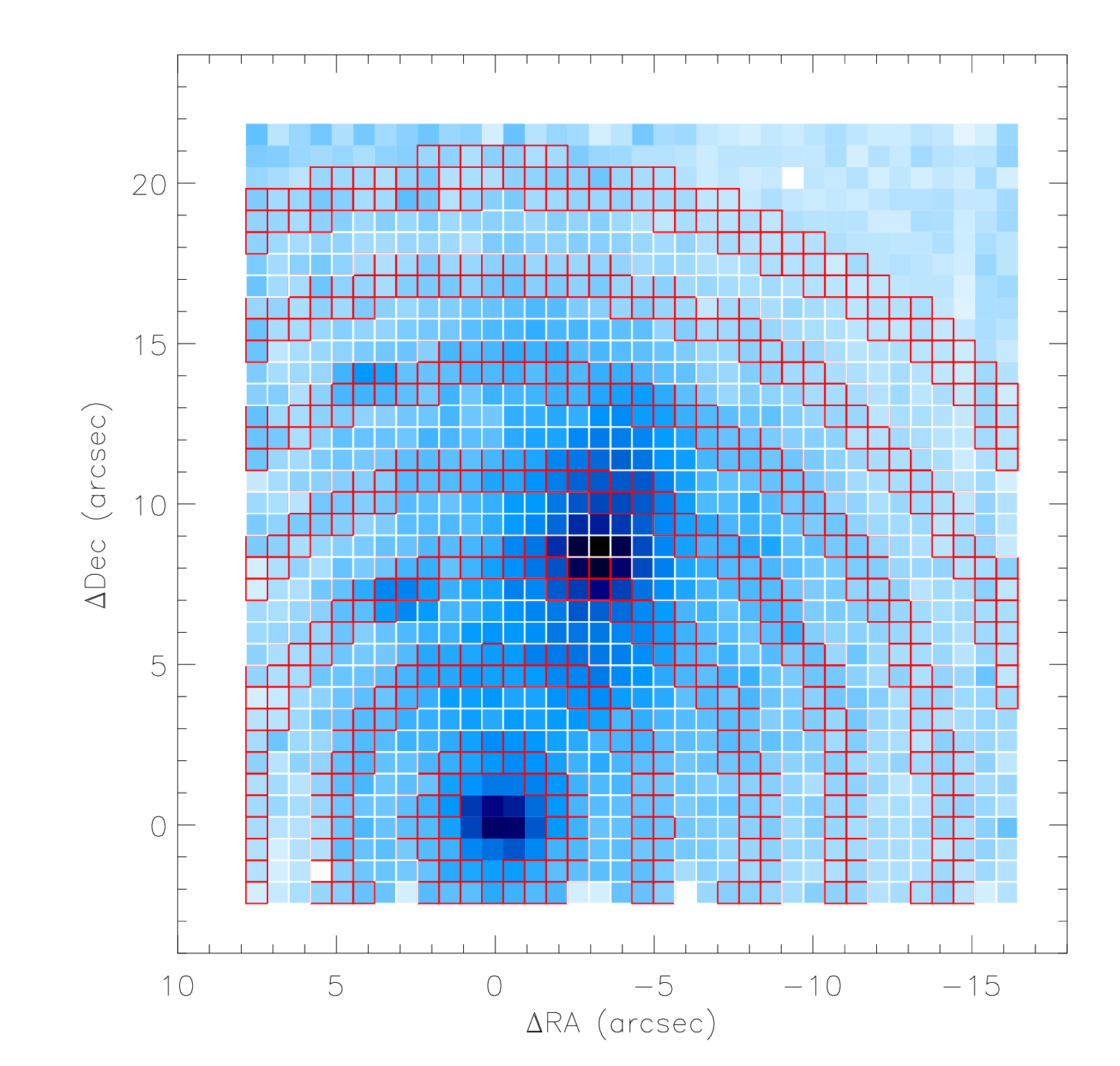}
    \includegraphics[height=6cm]{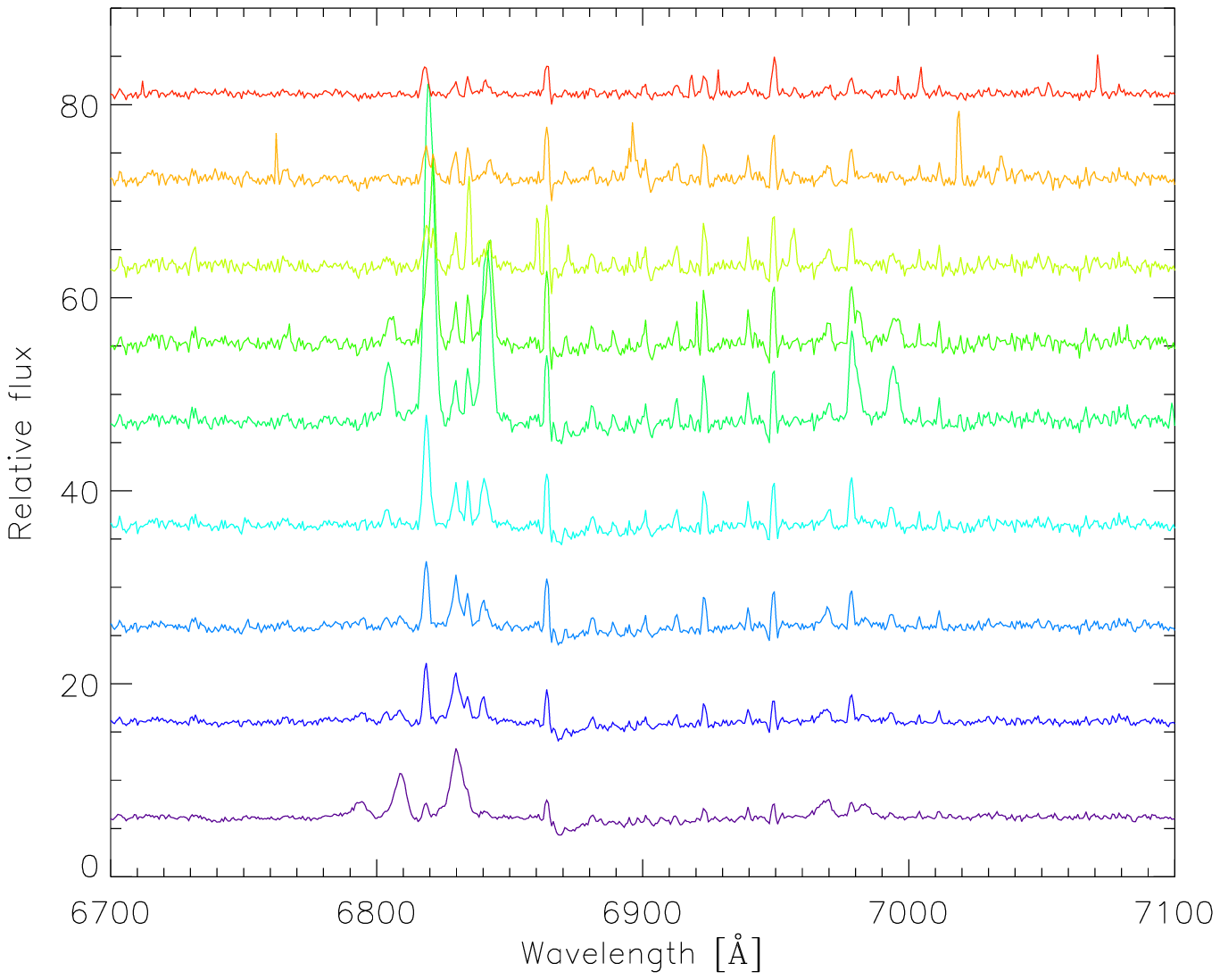}\hspace{0.2cm}
    \includegraphics[height=6cm]{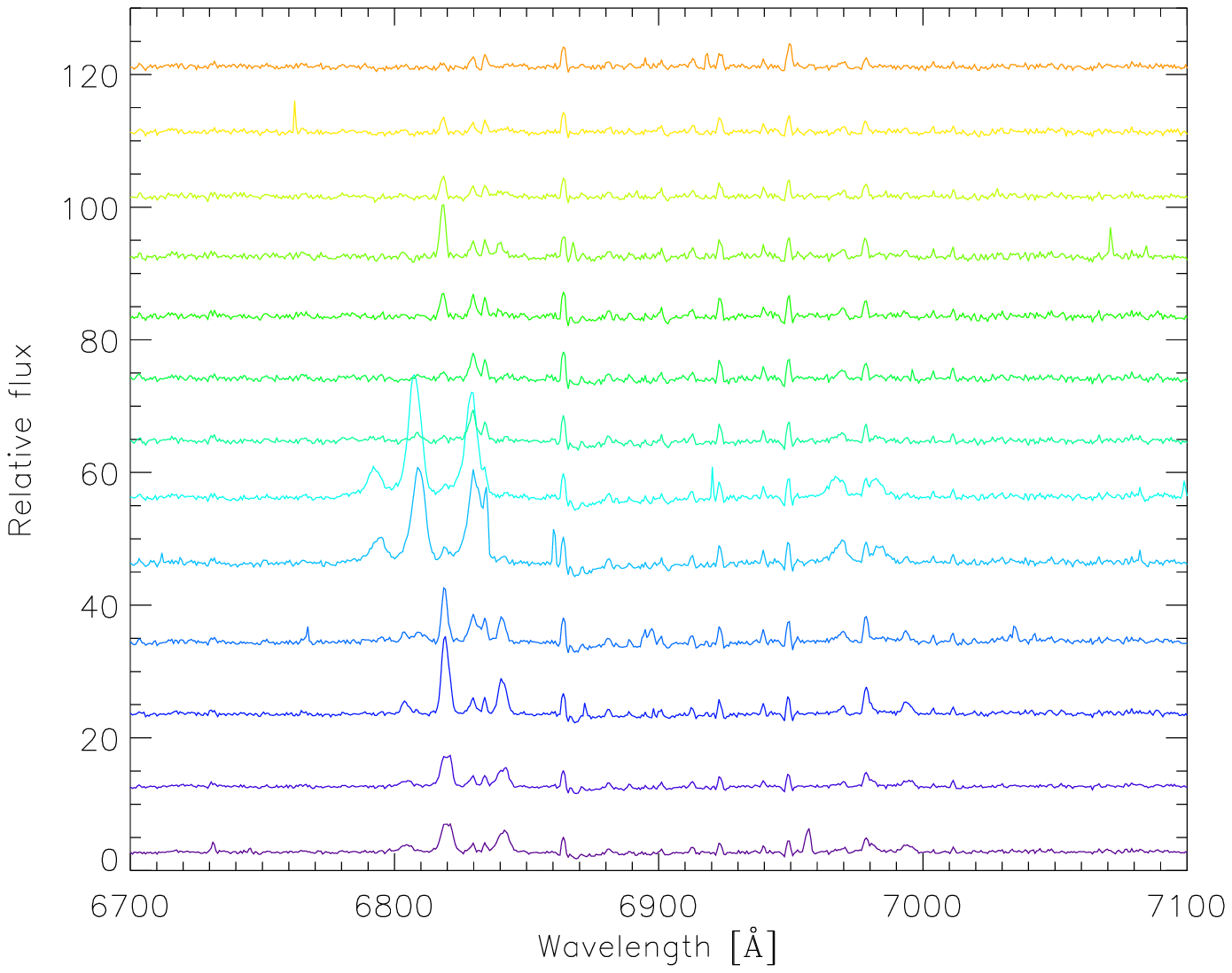}
    \caption[]{
      Example of an average radial extraction using both the 
      {\tt extract\_radial} and {\tt shift\_ptable} routines
      on a VIMOS field of the local LIRG \astrobj{IRAS F06076-2139}, after
      \citet{Arribas:2008p3550}. Top left-panel: extraction performed at the
      centre of the field. Top right-panel: extraction performed with a shifted
      reference point, centered on the nucleus of the South-East galaxy. The
      bottom panels show the integrated spectra within each ring, with radius
      increasing from the bottom to top in each case.
      \label{fig:radial}
    }
  \end{center}
\end{figure}

\noindent{\bf Spectra extraction and integration}
\vspace{0.4cm}

\noindent{\tt extract\_index}:
Extracts new a RSS file and generates a new position table based on an index
vector.
\vspace{0.2cm}

\noindent{\tt extract\_region}:
Extracts the spectra of regions selected by hand.
\vspace{0.2cm}

\noindent{\tt extract\_slit}:
Extracts the spectra within a rectangular aperture, resembling a long-slit
observation (see left-panel of \autoref{fig:extract}).
\vspace{0.2cm}

\noindent{\tt extract\_aperture}:
Extracts the spectra within a circular aperture (see right-panel of
\autoref{fig:extract}).
\vspace{0.2cm}

\noindent{\tt extract\_radial}:
Extracts radial average spectra within consecutive rings from a reference
point (see \autoref{fig:radial}).
\vspace{0.2cm}

\noindent{\tt integrate\_rss}:
Integrates the spectra contained in a RSS file into a single spectrum.
\vspace{0.8cm}

\noindent{\bf RSS and FITS manipulation}
\vspace{0.4cm}

\noindent{\tt read\_rss}:
Reads a RSS FITS file and stores the data into an IDL vector.
\vspace{0.2cm}

\noindent{\tt merge\_rss}:
Merges a list of RSS files into a single RSS file.
\vspace{0.2cm}

\noindent{\tt show\_hdr}:
Shows on screen the header of a FITS file, which can be written to an ASCII file.
\vspace{0.2cm}

\noindent{\tt write\_hdr}:
Adds or updates an entry in the header of a FITS file, using the
{\tt fxaddpar.pro} utility.
\vspace{0.2cm}

\noindent{\tt copy\_hdr}:
Copies the header of one FITS file to another.
\vspace{0.8cm}

\noindent{\bf Miscellaneous codes}
\vspace{0.4cm}

\noindent{\tt cube2rss}:
Converts a 3D FITS cube with dimensions $X$, $Y$, $\lambda$ to a RSS FITS file +
an {\em ad hoc} position table in ASCII format.
\vspace{0.2cm}

\noindent{\tt write\_wcs}:
Adds or updates the WCS (World Coordinate Systems) entries in a FITS header.
\vspace{0.2cm}

\noindent{\tt get\_new\_pt}:
Generates a new position table based on an index of selected spaxels.
\vspace{0.2cm}

\noindent{\tt shift\_ptable}:
Shifts the reference point or applies an offset to a given position table (an
application of this routine is shown in \autoref{fig:radial}).
\vspace{0.2cm}

\noindent{\tt merge\_ptable}:
Concatenates a list of position table files into a single one for mosaicking
purposes.
\vspace{0.2cm}

\noindent{\tt offset2radec}:
Transforms small angle offsets in arcsec from a reference point to equatorial
coordinates.
\vspace{0.2cm}

\noindent{\tt radec2offset}:
Transforms equatorial coordinates to small angle offsets from a given
reference point.
\vspace{0.2cm}

\section{Summary}

The {\sc PINGSoft} package presented in this article is a set of IDL routines
developed for the {\sc PINGS} project with a special emphasis on visualisation and
manipulation of IFS-based data. One of its major advantages with respect other
IFS visualisation tools reside in its portability to practically any OS platform
with a running version of the IDL data language, a common software in most
astronomical research institutes nowadays. The code is completely transparent to
the user, allowing to create tailored-based routines depending on each scientific
case. The package is run via command lines, which allows more flexibility during
repetitive tasks (especially when dealing with large databases), and more
precision in some cases (e.g. when defining extraction position/apertures). On
the other hand, the spatial and spectral visualisations are fully interactive
and optimised for large data files (e.g. mosaics), making the visualisation
rendering faster and less prone to errors than other GUI-based visualisation
tools.

{\sc PINGSoft} includes routines to extract regions of interest by hand or
within a given geometric aperture, to integrate the spectra within a given
region, to convert 3D cubes to the RSS format, to read, edit and write RSS FITS
files, and some other miscellaneous codes especially useful in astronomy and
spectroscopy. {\sc PINGSoft} is far from being perfect or complete, the main
intention is to help a broad audience to be more familiar with IFS data, but
bugs, errors and inconsistencies (especially with instruments not tested so
far) are expected. For comments, suggestions and bug reports please contact the
\href{mailto:frosales@cantab.net}{author}. The {\sc PINGSoft} package is freely
available at:\\

\noindent  \url{http://www.ast.cam.ac.uk/research/pings}\\

\noindent under the {\sf Software} section.
If you find this package useful on your research, please acknowledge the use
of {\sc PINGSoft} by citing the corresponding reference in your
publications. {\sc PINGSoft} is licensed under GPLv3\footnote{The GNU General
  Public License, found at: \url{http://www.gnu.org/licenses/gpl.html}}.

\vspace{0.5cm}
\noindent {\em Acknowledgements.} 
I would like to acknowledge the Mexican National Council for Science and
Technology (CONACYT) and the Direcci{\'o}n General de Relaciones Internacionales
(SEP) for the financial support during the period in which PINGSoft was
developed. 

\bibliographystyle{elsarticle-harv}
\bibliography{elsarticle}

\begin{thebibliography}{13}
\expandafter\ifx\csname natexlab\endcsname\relax\def\natexlab#1{#1}\fi
\expandafter\ifx\csname url\endcsname\relax
  \def\url#1{\texttt{#1}}\fi
\expandafter\ifx\csname urlprefix\endcsname\relax\def\urlprefix{URL }\fi

\bibitem[{Arribas et~al.(2008)Arribas, Colina, Monreal-Ibero, Alfonso,
  Garc{\'\i}a-Mar{\'\i}n, and Alonso-Herrero}]{Arribas:2008p3550}
Arribas, S., Colina, L., Monreal-Ibero, A., Alfonso, J.,
  Garc{\'\i}a-Mar{\'\i}n, M., Alonso-Herrero, A., Mar 2008. Vlt-vimos integral
  field spectroscopy of luminous and ultraluminous infrared galaxies. i. the
  sample and first results. A\&A 479, 687.

\bibitem[{Fevre et~al.(1998)Fevre, Vettolani, Maccagni, Mancini, Picat,
  Mellier, Mazure, Saisse, et~al.}]{LeFevre:1998p3053}
Fevre, O.~L., Vettolani, G.~P., Maccagni, D., Mancini, D., Picat, J.~P.,
  Mellier, Y., Mazure, A., Saisse, M., et~al., Jul 1998. Virmos: visible and
  infrared multiobject spectrographs for the vlt. Proc. SPIE Vol. 3355 3355, 8.

\bibitem[{Kelz and Roth(2006)}]{Kelz:2006p338}
Kelz, A., Roth, M.~M., Jun 2006. Experiences with the pmas-ifus. NewAR 50, 355.

\bibitem[{Kelz et~al.(2006)Kelz, Verheijen, Roth, Bauer, Becker, Paschke,
  Popow, S{\'a}nchez, et~al.}]{Kelz:2006p3341}
Kelz, A., Verheijen, M. A.~W., Roth, M.~M., Bauer, S.~M., Becker, T., Paschke,
  J., Popow, E., S{\'a}nchez, S.~F., et~al., Jan 2006. Pmas: The potsdam
  multi-aperture spectrophotometer. ii. the wide integral field unit ppak. PASP
  118, 129.

\bibitem[{Rosales-Ortega et~al.(2010)Rosales-Ortega, Kennicutt, S{\'a}nchez,
  D{\'\i}az, Pasquali, Johnson, and Hao}]{RosalesOrtega:2010p3794}
Rosales-Ortega, F.~F., Kennicutt, R.~C., S{\'a}nchez, S.~F., D{\'\i}az, A.~I.,
  Pasquali, A., Johnson, B.~D., Hao, C.~N., Mar 2010. Pings: the ppak ifs
  nearby galaxies survey. MNRAS, 461.

\bibitem[{Roth et~al.(2005)Roth, Kelz, Fechner, Hahn, Bauer, Becker, B{\"o}hm,
  Christensen, et~al.}]{Roth:2005p2463}
Roth, M.~M., Kelz, A., Fechner, T., Hahn, T., Bauer, S.-M., Becker, T.,
  B{\"o}hm, P., Christensen, L., et~al., Jun 2005. Pmas: The potsdam
  multi-aperture spectrophotometer. i. design, manufacture, and performance.
  PASP 117, 620.

\bibitem[{S{\'a}nchez(2004)}]{Sanchez:2004p2632}
S{\'a}nchez, S.~F., Mar 2004. E3d, the euro3d visualization tool i: Description
  of the program and its capabilities. AN 325, 167.

\bibitem[{S{\'a}nchez(2006)}]{Sanchez:2006p331}
S{\'a}nchez, S.~F., Jan 2006. Techniques for reducing fiber-fed and
  integral-field spectroscopy data: An implementation on r3d. AN 327, 850.

\bibitem[{Sandin et~al.(2010)Sandin, Becker, Roth, Gerssen, Monreal-Ibero,
  B{\"o}hm, and Weilbacher}]{Sandin:2010p3798}
Sandin, C., Becker, T., Roth, M.~M., Gerssen, J., Monreal-Ibero, A., B{\"o}hm,
  P., Weilbacher, P., Feb 2010. p3d: a general data-reduction tool for
  fiber-fed integral-field spectrographs. arXiv 1002, 4406.

\bibitem[{van~der Hulst et~al.(1992)van~der Hulst, Terlouw, Begeman, Zwitser,
  and Roelfsema}]{vanderHulst:1992p3831}
van~der Hulst, J.~M., Terlouw, J.~P., Begeman, K.~G., Zwitser, W., Roelfsema,
  P.~R., Jan 1992. The groningen image processing system, gipsy. Astronomical
  Data Analysis Software and Systems I 25, 131.

\bibitem[{Verheijen et~al.(2004)Verheijen, Bershady, Andersen, Swaters,
  Westfall, Kelz, and Roth}]{Verheijen:2004p2481}
Verheijen, M. A.~W., Bershady, M.~A., Andersen, D.~R., Swaters, R.~A.,
  Westfall, K., Kelz, A., Roth, M.~M., Mar 2004. The disk mass project; science
  case for a new pmas ifu module. AN 325, 151.

\bibitem[{Walsh and Roth(2002)}]{Walsh:2002p3819}
Walsh, J.~R., Roth, M.~M., Sep 2002. Developing 3d spectroscopy in europe. The
  Messenger (ISSN 0722-6691) 109, 54.

\bibitem[{Westmoquette et~al.(2009)Westmoquette, Exter, Christensen, Maier,
  Lemoine-Busserolle, Turner, and Marquart}]{Westmoquette:2009p3557}
Westmoquette, M.~S., Exter, K.~M., Christensen, L., Maier, M.,
  Lemoine-Busserolle, M., Turner, J., Marquart, T., May 2009. The integral
  field spectroscopy (ifs) wiki. arXiv 0905, 3054.

\end{thebibliography}

\end{document}